\RequirePackage{fix-cm}
\documentclass[twocolumn]{svjour3}

\usepackage{amsmath}
\usepackage{amssymb}
\usepackage[dvips]{graphicx}

\def\ve{\varepsilon}

\journalname{Granular Matter}

\begin{document}

\title{A contact model for the yielding of caked granular materials}

\author{L.\ Brendel \and J.\ T\"or\"ok \and R.\ Kirsch \and U.\ Br\"ockel}

\institute{
L.\ Brendel, J.\ T\"or\"ok \at
Faculty of Physics, University of Duisburg-Essen, 47048 Duisburg, Germany\\
\email{lothar.brendel@uni-due.de}
\and
R.\ Kirsch, U.\ Br\"ockel \at
Fachhochschule Trier, Umwelt-Campus Birkenfeld 
55761 Birkenfeld, Germany
}

\date{Received: date / Accepted: date}

\maketitle

\begin{abstract}
  We present a visco-elastic coupling model between caked spheres,
  suitable for DEM simulations, which incorporates the different
  loading mechanisms (tension, shear, bending, torsion) in a combined
  manner and allows for a derivation of elastic and failure properties
  on a common basis. In pull, shear, and torsion failure tests with
  agglomerates of up to 10000 particles, we compare the failure
  criterion to different approximative variants of it, with respect to
  accuracy and computational cost. The failure of the agglomerates,
  which behave according to elastic parameters derived from the
  contact elasticity, gives also insight into the relative relevance
  of the different load modes.
\end{abstract}


\section{Introduction}
\label{Sec_int}

The research of granular media can be separated into two rather
distinct fields: There are systems with purely repulsive granular
interaction (``dry granular media'', e.g.\ \cite{PhysDryGran_97})
and media with non-negligible additional attractions (``wet granular
media'', e.g.\ \cite{Mitarai_Nori_06}). While for the former the further
distinction between granular gases (e.g.\ \cite{GranGas_01}) and dense
systems can be made, the latter is more adequately separated into interactions of
permanent or recurring nature (like e.g.\ van der Waals
attraction or liquid menisci) and those which vanish irreversibly when
being overcome (like sinter necks or other solid bridges). In this
work, we focus solely on this last type.

Such cemented but breakable contacts form (deliberately or accidentally) in
agglomerates of particles. The failure behavior of these agglomerates
is of practical interest in both cases, but
also from the point of view of Distinct Element Methods (DEMs),
modelling the hindering of all six relative degrees of freedom of two
particles in contact is much less studied than the well established
frictional contact models like the linear
spring-dash pot, the Hertz contact and its extensions (cf.\
e.g.~\cite{Roux_in_PhysDryGran,Luding_in_PhysDryGran}).

For modelling purposes two aspects of
the contacts have to be taken into account: its elastic response to
deformations and the maximal load it can bear. The elastic properties
are usually expressed as set of (often decoupled) springs associated
to each relative degree of motion (cf.\
e.g.~\cite{Delenne_etal_04,Antonyuk_etal_06,Wang_AMarroquin_09}). Also
for the maximal load it is a common practice to employ separate
thresholds. Exceptions (in the context of granular assemblies)
to an independent treatment of just tensile and shear
load are \cite{Cundall_Strack_79,Delenne_etal_04,Wang_AMarroquin_09}
who employ empirical or plausible combination functions.

It is thus legitimate to use a simple but self consistent method
to model the cemented contact by a flat, elastic cylinder subjected to
the Tresca failure criterion (section \ref{Sec_model}). The full
Tresca criterion is only solvable with iterations therefore we
compare the full criterion to its two simplified variants in DEM
simulations of the quasi-brittle failure of a cylinder made up of 200
to 10000 particles (section \ref{Sec_sim}). We conclude with
discussing the relative relevance of the different loading modes in
section \ref{Sec_failuretype}.


\section{Modelling breakable cemented contacts}
\label{Sec_model}

\subsection{Contact model}
\label{Sec_contmodel}

The principle modelling idea for the cemented contact between two
particles is the dominance of the bridge with respect to the elastic
and failure properties. Hence, we envisage the bridge as a flat
cylinder (height $d$ smaller than diameter $2a$) while treating the
rest of the two particles as solid bodies.

We set up a local coordinate frame, where the bottom ``plate'' of the
cylinder is fixed in the $x/y$-plane with its center in the origin,
while the upper one is located (in the unstrained state) parallel at
$z=d$ (i.e.\ the original contact normal is in the $z$ direction). We
call the torsion about the $z$-axis ``twist'' with angle $\phi$, while
we choose the $x$-axis of the local frame as the axis of the other
rotational load, the ``tilt'' $\theta$ of the upper plate. These
rotations (being, due to the brittle limit, small as to allow for
commutation) are accompanied by the translational displacements, the
pull/push $Z$ and the shear $X,~Y$. In total, we describe the load
geometrically by these five parameters. This is independent of the
actual representation of particle orientation e.g.\ in a DEM code.

Next, we need to know the elastic response in terms of pull/push force
$F_z$, shear forces $F_x, F_y$, and torques $T_z, T_x$ opposing the
twist and tilt, respectively. We avoid introducing five arbitrary
spring constants, and aim to derive them from the elastic behavior of
the cylinder, instead.

The stress inside a deformed cylinder (possessing elastic modulus $E$
and Poisson ratio $\nu$) is in principle a text-book problem (cf.\
e.g.~\cite{Sokolnikoff_book}). We directly provide the displacement
field associated to each loading mode:
\begin{align}
  \label{eq_displacement1}
  \vec u_\text{push/pull}(x,y,z) &=\frac{F_z}{E \pi a^2}\left(-\nu x\vec
    e_x-\nu x\vec e_y+z\vec e_z\right)\\
  \vec u_\text{torsion}(x,y,z) &= \frac{4 T_z(1+\nu)}{E \pi a^4}\left(-yz\vec
    e_x+xz\vec e_y\right)\\
  \notag
  \vec u_\text{bend}(x,y,z) &= \frac{2 T_x}{E \pi a^4}\left(-2\nu xy\vec e_x
    +2\nu yz\vec e_z\right.\\
  &\quad\left.-(\nu(y^2-x^2)+z^2)\vec e_y\right)\\
  \label{eq_displacement4}
  \vec u_\text{shear}(x,y,z) &= \frac{2+2\nu}{E\pi a^2}\left(F_x z\vec
    e_x+F_y z\vec e_y\right)
\end{align}
From these fields, we can immediately read off the deformation geometry as
\begin{align}
  \label{eq_deform}
  X&=\vec e_x\cdot\vec u_\text{shear}(0,0,d)=\frac{2d(1+\nu)}{E\pi a^2}F_x\\
  Y&=\vec e_y\cdot\vec u_\text{shear}(0,0,d)=\frac{2d(1+\nu)}{E\pi a^2}F_y\\
  Z&=\vec e_z\cdot\vec u_\text{push/pull}(0,0,d)=\frac{d}{E\pi a^2}F_z\\
  \phi&=\vec e_y\cdot \vec u_\text{torsion}(a,0,d)/a=\frac{4d(1+\nu)}{E \pi a^4}T_z\\
  \theta&=-\partial_z (\vec e_y\cdot\vec u_\text{bend}(0,0,d))=\frac{4d}{E
    \pi a^4}T_x
  \quad.
\end{align}

Summing up (\ref{eq_displacement1}) to (\ref{eq_displacement4}),
calculating the corresponding strain and applying Hook's law yields a
stress tensor of
\begin{align}
  \label{eq_stresstensor1}
  \sigma_{xx}&=\sigma_{xy}=\sigma_{xy}=\sigma_{yy}=0\\
  \sigma_{xz}&=\sigma_{zx}=\frac{a^2 F_x-2T_z y}{\pi a^4}\\
  \sigma_{yz}&=\sigma_{zy}=\frac{a^2 F_y+2T_z x}{\pi a^4}\\
  \label{eq_stresstensor4}
  \sigma_{zz}&=\frac{a^2 F_x+4T_x y}{\pi a^4}
\end{align}
which obeys vanishing divergence and vanishing normal stress at the
free surface. Integration over the lateral surfaces confirms the force
$\vec F=F_x\vec e_x+F_y\vec e_y+F_z\vec e_z$ and reveals a torque
(width respect to the point $\vec e_z d/2$) of $\vec T=(T_x+F_x
d/2)\vec e_x+F_y d/2\vec e_y+T_z\vec e_z$ on the bottom one (and thus
$-\vec F$ and $-\vec T$ on the top one, of course). The additional
torque contributions are indeed necessary, since we defined the shear
mode as keeping bottom and top plate parallel.

With the definition of the spring constant $k_n\equiv E \pi a^2/d$
(which in principle can be measured in experiments with two particles
\cite{Birkenfeld}), the relation between deformation and reaction
force finally reads
\begin{equation}
  \label{eq_mstiff}
  \begin{pmatrix}
    F_x\\ F_y\\ F_z\\ T_x/a\\ T_y/a\\ T_z/a
  \end{pmatrix}
  =-k_n
  \begin{pmatrix}
    \frac{1}{2+2\nu} & 0 & 0 & 0 & 0\\
    0 & \frac{1}{2+2\nu} & 0 & 0 & 0\\
    0 & 0 & 1 & 0 & 0\\
    \frac{1}{2+2\nu} \frac{d}{2a} & 0 & 0 & \frac{1}{4} & 0\\
    0 & \frac{1}{2+2\nu} \frac{d}{2a} & 0 & 0 & 0\\
    0 & 0 & 0 & 0 & \frac{1}{4+4\nu}
  \end{pmatrix}
  \begin{pmatrix}
    X\\ Y\\ Z\\ a\theta\\ a\phi
  \end{pmatrix}
  \quad.
\end{equation}
When considering the cylinder (the geometry of which does not
explicitly enter into the contact geometry anyway) as being very flat
(i.e.\ $d/a\to 0$), the stiffness matrix looses the $T_y$-line and
becomes diagonal. This decoupling is very convenient for DEM
modelling. Moreover, every force/torque component can be supplemented
by a viscous damping involving the time derivative of the deformation
and with coefficients for critical damping:
\begin{align}
  \label{eq_diagndamp}
  -F_x &=2\sqrt{k_t\bar m}\,\dot X+k_t\,X\\
  -F_y &=2\sqrt{k_t\bar m}\,\dot Y+k_t\,Y\\
  -F_z &=2\sqrt{k_n\bar m}\,\dot Z+k_n\,Z\\
  -T_x &=a\sqrt{k_n\bar I_x}\,\dot\theta+\frac{k_na^2}{4}\,\theta\\
  -T_z &=a\sqrt{2k_t\bar I_z}\,\dot\phi+\frac{k_ta^2}{2}\,\phi
\end{align}
Here, $k_t\equiv k_n/(2+2\nu)$ while $\bar m,~\bar I_x$, and $\bar
I_z$ are the reduced mass and moments of inertia of the two-particle
system, respectively. Critical damping is used for convenience when
the deformation rate is anyway slow enough as to neglect velocity
effects.


\subsection{Failure criterion}
\label{Sec_breakcrit}

Having worked out the elastic response of the bridge, we now turn to
its failure. The choice of failure criteria for condensed matter to be
found in the literature is vast. We will use the well known Tresca
criterion\cite{Tresca_1864} of maximal shear stress. From the stress tensor
(\ref{eq_stresstensor1})-(\ref{eq_stresstensor4}) the principle
stresses are readily found to be
\begin{align}
  \label{eq_prstresses}
  \pi a^4\sigma_1&=a^2F_z/2+2T_x y-a^2\sqrt{\Phi^2}\\
  \pi a^4\sigma_2&=0\\
  \pi a^4\sigma_3&=a^2F_z/2+2T_x y+a^2\sqrt{\Phi^2}\\
  \intertext{where}
  \notag
  a^4\Phi^2 &= a^4(F_x^2+F_y^2+F_z^2/4)+4(T_x^2 y^2+T_z^2(x^2+y^2))\\
  &+2a^2(F_z T_x y+2T_z(F_y x-F_x y))
  \quad.
\end{align}
Since $\sigma_1<\sigma_2<\sigma_3$\footnote{We use negative sign for
  compressive stress.}, the maximal shear stress (with respect to
orientation) is
\begin{equation}
  \label{eq_tresca}
  \sigma_\text{Tresca}(x,y)=\frac{\sigma_3-\sigma_1}{2}=\frac{\sqrt{\Phi^2}}{\pi a^2}
\end{equation}
and has yet to be maximized with respect to $x^2+y^2\le a^2$ and then
compared to a material dependent critical stress $\sigma_*$. That
means, we are left with a maximization of $\Phi^2$. Introducing forces
$\bar T_x=T_x/a$, $\bar T_z=T_z/a$ and dimensionless position $\bar
x=x/a$, $\bar y=y/a$ yields
\begin{equation}
  \label{eq_Phi2}
  \bar \Phi^2=\bar T_z^2\left(\bar x+\frac{F_y}{2\bar T_z}\right)^2
  +(\bar T_x^2+\bar T_z^2)\left(
    \bar y+\frac{F_z\bar T_x-2F_x\bar T_z}{4(\bar T_x^2+\bar T_z^2)}
  \right)^2
\end{equation}
where $\Phi^2$ and $4\bar \Phi^2$ differ only by an irrelevant $\bar x/\bar
y$-independent term. Obviously, $\bar \Phi^2$ is just a quadratic form in
$\bar x$ and $\bar y$ the center and steepness of which depend on the
load. Unfortunately, when maximizing $\bar \Phi^2$ subjected to the
constraint $\bar x^2+\bar y^2\le 1$ (which, due to the absence of
local maxima, can be replaced by $\bar x^2+\bar y^2=1$) one is faced
with a quartic, the radicals of which exceed any sensible formula
length.

For several special cases, the calculation can be performed, though:
In the simplest case of zero torques, $\sigma_\text{Tresca}$ becomes
independent of $x$ and $y$ and thus
\begin{equation}
  \label{eq_notorques}
  \pi a^2 \sigma_\text{Tresca,max}\equiv F_\text{Tresca}=\sqrt{F_x^2+F_y^2+F_z^2/4}
\end{equation}
and can be compared directly to the threshold $F_*\equiv\pi
a^2\sigma_*$.
A less simple case is $F_y=0$ which lets $\bar \Phi^2$ take on its
maximum at $\bar y=\mathrm{sgn}(F_z T_x-2F_xT_z)$, producing the
value
\begin{equation}
  \label{eq_noFy}
  F_\text{Tresca}=\sqrt{F_x^2+\frac{F_z^2}{4}+2\frac{\lvert
      F_zT_x-2F_xT_z\rvert}{a}
    +4\frac{T_x^2+T_z^2}{a^2}}
  \quad.
\end{equation}
Yet another illuminating combination, namely $F_x=F_z=0$, yields
\begin{equation}
  \label{eq_Phi2noFxnoFz}
  \bar \Phi^2=\frac{(F_y+2\bar T_z\bar x)^2}{4}+(\bar T_x^2+\bar
  T_z^2)\bar y^2
\end{equation}
which possesses a local maximum at $\bar x=F_y T_z/(2T_x^2)$, $\bar
y^2=1-\bar x^2$ (the sign of $\bar y$ being irrelevant). This is valid
only for $\lvert F_y \bar T_z\rvert<2\bar T_x^2$, otherwise $\bar
x=\mathrm{sgn}(F_y\bar T_z),~\bar y=0$ has to be chosen. In the latter
case, the value taken on is
\begin{equation}
  \label{eq_noFxnoFz}
  F_\text{Tresca}=\sqrt{F_y^2+4\frac{\lvert F_y T_z\rvert}{a}+4\frac{T_z^2}{a^2}}
\end{equation}

For the general case (occurring naturally in simulations), we
resorted to locating the maximum of (\ref{eq_Phi2}) 
numerically (by means of a Lagrangian multiplier, the Newton-Raphson
method and being careful about poles). Let us term this and the
subsequent usage of the maximized $\Phi$ in (\ref{eq_tresca}) the
application of the \emph{full (Tresca-based) criterion}. In section
\ref{Sec_yieldtests}, we will come back to it and compare the results
to the ones obtained by employing the following rather crude estimate.

To obtain an approximate closed formula,
we disregard the exact constraint $\bar x^2+\bar y^2=1$ and replace it
by $\bar x^2=\bar y^2=1$. Then again, the lack of local maxima in
$\bar \Phi^2$ enforces its maximum to be located at one of the
four corner points $(\bar x,\bar y)=(\pm 1,\pm 1)$. This in turn, given the
form of (\ref{eq_Phi2}), shows that the signs of $\bar x$ and $\bar y$
have to be those of $F_y/\bar T_z$ and $F_z\bar T_x-2F_x\bar T_z$,
respectively. Plugging that back into (\ref{eq_tresca}) yields
\begin{align}
  \notag
  \tilde F^2_\text{Tresca}&=F_x^2+F_y^2+F_z^2/4+4(T_x^2+2T_z^2)/a^2\\
  \label{eq_crude}
  &+(2\lvert F_zT_x-2F_xT_z\rvert+4\lvert F_y T_z\rvert)/a
  \quad.
\end{align}
Comparing it to the special cases (\ref{eq_notorques}),
(\ref{eq_noFy}), and (\ref{eq_noFxnoFz}), we recognize that the
estimation can be improved by just dropping the factor $2$ in front of
$T_z^2$, i.e.\ we have to check
\begin{align}
  \notag
  F^2_\text{Tresca}&=F_x^2+F_y^2+\frac{F_z^2}{4}+4\frac{T_x^2+T_z^2}{a^2}\\
  \label{eq_lesscrude}
  &+\frac{2\lvert F_zT_x-2F_xT_z\rvert+4\lvert F_y T_z\rvert}{a}
\end{align}
against the threshold $F_*^2$. Let us call this the \emph{simplified
criterion}.

For comparison purposes, we introduce the \emph{decoupled criterion} as well,
where all loading modes are checked individually:
\begin{equation}
  \label{eq_decoupled}
  F_\text{Tresca}=\max\left\{\sqrt{F_x^2+F_y^2},\lvert
    F_z\rvert/2,2\lvert T_x\rvert/a,2\lvert T_z\rvert/a\right\}
\end{equation}

Whatever criterion (i.e.\ definition of $F_\text{Tresca}$) used, in
the case of exceeding the threshold (i.e.\ $F_\text{Tresca}>F_*$), the
contact is irreversibly transformed into a usual, frictional linear
spring-dash pot one with the same visco-elastic parameters.


\section{Simulation}
\label{Sec_sim}

\subsection{Simulation setup}
\label{Sec_simsetup}

In this part, we describe in detail the procedure which was used to
create standard yield tests on cylindrical shaped specimens using the
LAMMPS \cite{LAMMPS_95} molecular dynamics simulation code, extended
to treat caked contacts according to the model in section \ref{Sec_model}.

In order to create a homogeneous, random close packing, first
particles were poured into a half cylinder placed horizontally in a
vertical gravity field. The cylinder had the same diameter as the
desired test specimen but was necessarily much longer.

Particles had a narrow size distribution that is large enough to
prohibit crystallization on the cylinder wall. The number of particles
varied from $200$ to $10000$, the friction coefficient in the
preparation as well as in the tests was set to $0$.

The radius of the cylinder was chosen to be
\begin{equation}\label{Eq_radius}
R=\left(\frac{N}{f\pi s}\right)^{1/3}\bar d,
\end{equation}
where $\bar d$ is the average particle diameter, $N$ is the number of
particles and $s$ is the desired height width sample ratio:
$s{=}H/(2R)$, for which values in the range of $0.5{-}10$ were
chosen. The value of $f$ describes the average volume taken up by a
particle as $2\bar d^3/f$ and thus, for a random close packing with a
volume fraction of around $0.65$, amounts to $f\approx 2.5$.

After all particles were poured into the cylinder the sample was
uni-axially compressed by two plates in absence of gravity. The
compression force employed was small enough to keep the average
deformations of the friction-less, linear spring-dash pot contacts
below $10^{-5}$ particle radii. For the subsequent failure tests, the
stiffness was reduced by one order of magnitude and the pairwise
distance was rescaled to avoid pre-stressing.

\begin{figure}
\includegraphics[width=0.32\columnwidth]{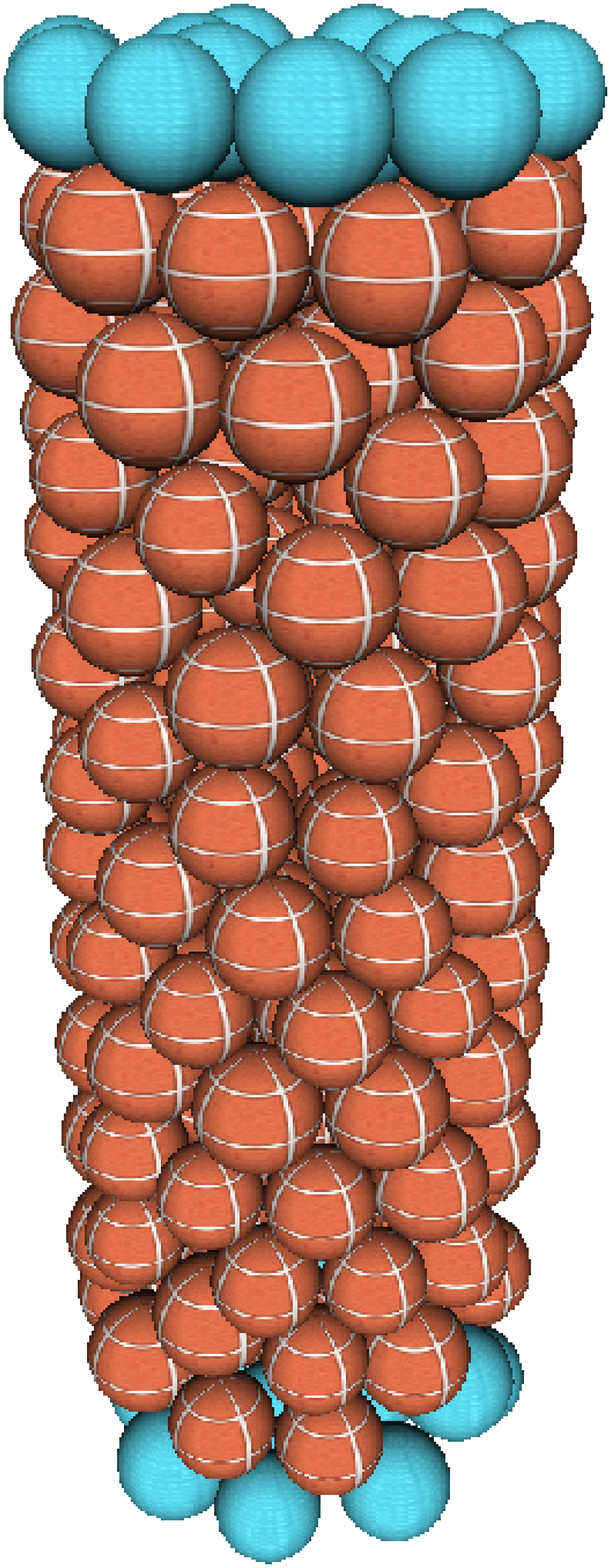}
\includegraphics[width=0.32\columnwidth]{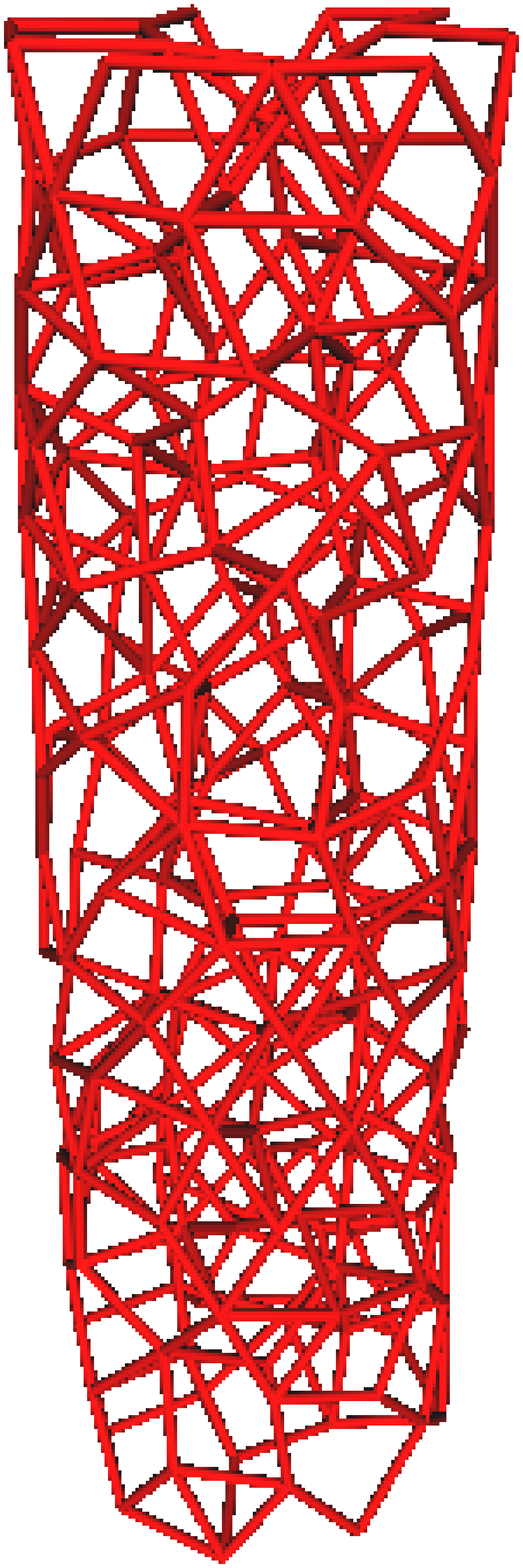}
\includegraphics[width=0.32\columnwidth]{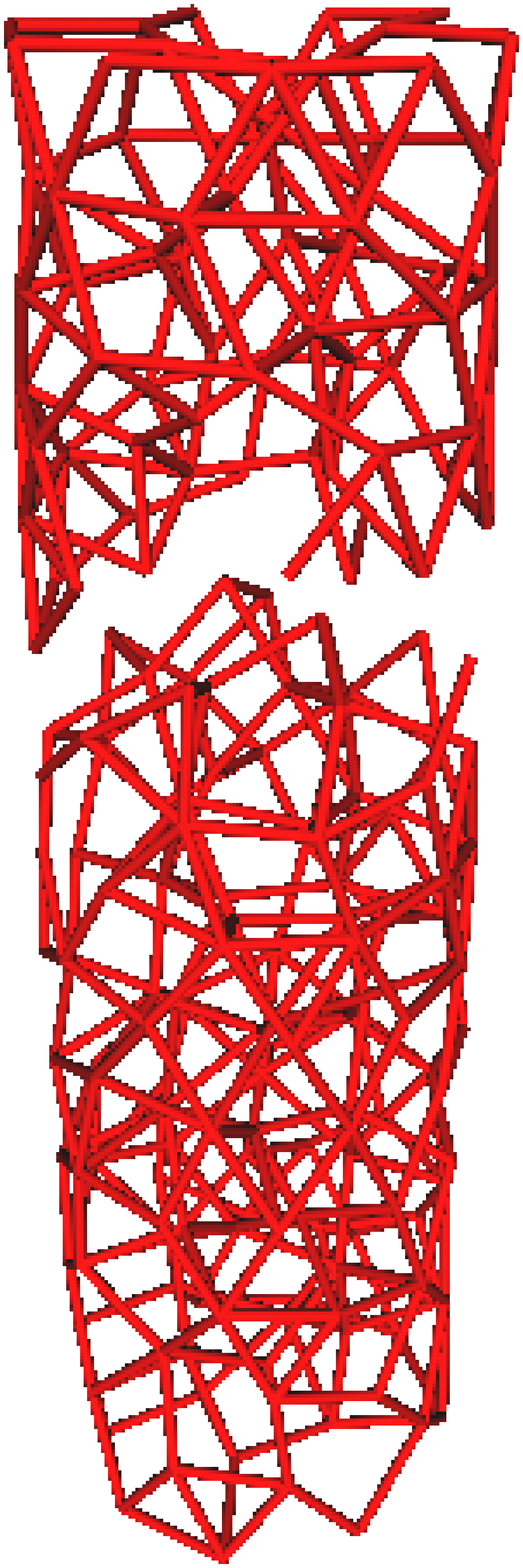}

{\null\hfill(a)\hfill\hfill(b)\hfill\hfill(c)\hfill\null}
\caption{\label{Fig_setup} (a) A typical test setup with aspect ratio
$s\simeq3$ and $N=200$. The smooth, blue particles are the driving
particles, the striped brown ones are the bulk particles. (b) The
contact network of the same system. (c) The broken contact network
after strain $\ve=0.00017$. Note, that particle displacement would not
be visible.}
\end{figure}

The tests were done in absence of gravity and without the compressing
plates. There are fewer contacts from the bulk to a plate than to a
layer of particles, thus if plates were used for driving it would be
expected that the contacts between a plate and the touching particles
would fail first. So we chose the last layers of particles (the ones
that are closer than $1.1$ particle radius to the compressing plate)
to move as a rigid body and drive the system. This allows the failure
zone to be everywhere in the sample. A typical setup and test result
is shown on Fig.~\ref{Fig_setup}.

Two different driving methods were used: (i) the driving layer had
infinite mass, (ii) the driving layer had infinite mass only in the
driving direction but finite in the perpendicular directions. Since no
significant difference was observed, the results reported in this
paper are shown with driving particles of infinite mass. The driving
velocity was always kept low enough as to avoid kinetic effects.

After the preparation the sample was tested. At the beginning of the
tests all existing (i.e.\ force transmitting) contacts were ``glued'',
all subsequent contacts (including broken ones at later times) were
handled with the standard Hooke contact law including static
friction. The simulation was run long enough such that the upper and
lower part are completely separated as shown on
Fig.~\ref{Fig_setup}(c).

The different tests were realized by giving the driving particles on
both sides of the sample initially a velocity (or angular velocity) in
the required direction but with opposing sign.

\subsection{Parameters}
\label{Sec_parameters}

In order to interpret the results of numerical simulations we have to
take a closer look at the dimensions of the parameters. Due to the
quasi static nature of the tests (up to the failure point) and the absence
of gravity, quantities like time and mass are unimportant, instead
{\em force} and {\em distance} units determine the scales of the
numerical results.

Two parameters of dimension \emph{distance} enter: the average
particle diameter $\bar d$ and the contact glue diameter $2a$. Since
in all simulations the ratio $a{=}0.1\bar d$ was used, $\bar d$ is
taken as the unit of length.

The stress or force scale is set by the stress threshold $\sigma_*$
(cf.~(\ref{eq_tresca})) and $F_*=\pi a^2\sigma_*$, respectively. Due to
technical reasons, 17 times the latter was chosen as unit of
force. 

Thus, from now on all quantities will be given in this natural units ($\bar
d$ for length and $17 F_*$ for force).

The normal contact stiffness $k_n$ (the tangential one is set
to $k_t{\equiv}k_n/2$, corresponding to the limit $\nu=0$,
cf.~(\ref{eq_mstiff})) relates force and length and must fulfill
$k_n\gg 1$ to assure the brittle limit. This can also be expressed as the
failure strain scale
\begin{equation}\label{Eq_maxstrainfromk}
  \ve_*=\frac{F_*}{k_n \bar d}
\end{equation}
being very small.

The preparation was done with the aim to achieve a sample as
homogeneous and isotropic as possible. For such a random
configuration, the macroscopic Young's modulus can be calculated
analytically \cite{micromacro2,micromacro1} from the stiffness
parameters of the contacts:
\begin{equation}
E=\frac{N_c\bar d}{3V}\frac{\left(2k_n+3k_t\right)}{4k_n+k_t}
\end{equation}
where $N_c$ is the number of contacts and $V$ is the volume of the
sample. The latter can be calculated using (\ref{Eq_radius}):
\begin{equation}\label{Eq_youngfromk}
E= \frac{\bar z f(2k_n+3k_t)}{12\bar d(4k_n+k_t)},
\end{equation}
where $\bar z$ is the average coordination number which was measured
to be $\bar z{\simeq}5.06$ in all cases. This value is less than the
isostatic limit for homogeneous packing due to the boundary effects.
Using the values $f{=}2.5$ and $k_t{=}k_n/2$, we get an estimate for
the macroscopic Young's modulus:
\begin{equation}\label{Eq_youngsestimate}
E= 0.81 \frac{k_n}{\bar d}
\end{equation}
The macroscopic Poisson ratio can also be expressed as function of the
contact stiffness \cite{micromacro1,micromacro2}:
\begin{equation}\label{Eq_poissonfromk}
\nu=\frac{k_n-k_t}{4k_n+k_t}=0.11
\end{equation}

The actual simulation parameters (in natural units) chosen for the
current study were taken to allow reliable and fast simulation of the
tests and yet fulfill the requirements above:
\begin{align}
\bar d &= 1 & a &= 0.1\cr
k_n &= 1256 & k_t &= k_n/2\cr
\label{Eq_params}
\sigma^* &= 1.88 & &~&
\end{align}

From those we can calculate the macroscopic Young's modulus
(\ref{Eq_youngfromk}) and Poisson ratio (\ref{Eq_poissonfromk}) as
well as the limit strain (\ref{Eq_maxstrainfromk}):
\begin{equation}\label{Eq_predict}
E = 1017
\quad,\quad
\nu = 0.11
\quad,\quad
\ve_* = 9.4\times10^{-5}
\end{equation}


\section{Yield tests}
\label{Sec_yieldtests}

\subsection{Poisson ratio}
\label{Sec_Poissonratio}

The easiest quantity that we can measure in our simulations is the
Poisson ratio. In pull experiments we recorded the displacement of the
driving particles in the $z$ direction and the radius of the sample
at $z{=}H/2$. The latter was done by averaging the outer radius
over all surface particles that have at least two contacts (remove
rattling, dangling particles). Since the relative displacement is very
small (of the order of $10^{-6}$) we can use the linearized expression
for the Poisson ratio:
\begin{equation}
\nu\simeq -\frac{\Delta R}{\Delta H}
\end{equation}
We averaged for different samples and system size to get the result:
\begin{equation}
\nu=0.108\pm0.01
\end{equation}
This value is in perfect agreement with the prediction of (\ref{Eq_predict}).

\subsection{Stress/strain behavior}
\label{Sec_stress_strain}

Three different tests were done: pull, shear and twist. In the tests
the driving was realized by giving the driving particles a constant
(angular) velocity. The net force (torque), the number of intact
contacts, the deformation of the sample was recorded as function of
the displacement of the driving particles.

The stress/strain evolution of the samples in the homogeneous and
isotropic case are described by three parameters: Young's
modulus\footnote{Just like in the previous section, we use the symbols
  $E$ and $\nu$ for the macroscopic quantities instead of the ones at
  contact level as in section \ref{Sec_model}. The same applies to the
  deformations $X,~Z$, to forces/torque $F_x,~F_z,~T$ and stresses.} $E$,
Poisson ratio $\nu$ and maximal strain $\ve^*$. So the evolution of
all test should be described by one single curve. Thus the measured
displacement and force (torque) data has to be converted to a
stress/strain relation.

It is more informative if this conversion is done with the
dimensionless quantities of the aspect ratio $s$ and the number of
particles $N$ since it allows immediately the scaling of different
tests in a transparent manner.

In general the stress and the strain are not homogeneous, so we focus
on their two important characteristics: the maximum and the average. 
The maximum stress governs the point where the first
contact breaks, the average value indicates the importance of the
inhomogeneities of the stress and we compare it to the maxima of the
stress/strain curve.

The pull deformation is homogeneous thus the maximum is the same as
the average:
\begin{equation}
\bar\ve_{p}=\ve_{p,\mathrm{max}}=
\frac{Z}{H}=\frac{(f\pi)^{1/3}Z}{2s^{2/3}N^{1/3}\bar d},
\end{equation}
where $Z$ is the vertical displacement of the driving particles and
$f$ is the volume factor of (\ref{Eq_radius}). The stress can be
taken directly from section \ref{Sec_contmodel}:
\begin{equation}
\bar\sigma_p=\sigma_{p,\mathrm{max}}=\frac{F_z}{\pi R^2}=
\left(\frac{fs}{\sqrt{\pi}N}\right)^{2/3}\frac{F_z}{\bar d^2}
\end{equation}

The twist deformation has a radius dependence, but apart from this is
very similar to the previous case.
\begin{align}
\ve_{t,\mathrm{max}}&=\frac{R\alpha}{H}=\frac{\alpha}{2s}\cr
\bar\ve_t&=2 \ve_{t,\mathrm{max}}/3,
\end{align}
where $\alpha$ is the angle covered by the driving layer. The form of
the stress again is identical to the microscopic one in
\ref{Sec_contmodel}:
\begin{align}
\sigma_{t,\mathrm{max}}&=\frac{4T(1+\nu)}{\pi R^3}=
\frac{4f(1+\nu)s}{N}\frac{T}{\bar d^3}\cr
\bar\sigma_t&=2 \sigma_{t,\mathrm{max}}/3
\end{align}

The shear deformation is a bit more involved than the previous ones,
since its behavior is fundamentally different whether the aspect ratio
is large or small compared to unity (with a crossover in between). The
short cylinder limit was covered in Sec.~\ref{Sec_contmodel}. The long
cylinder limit is more complicated, but consists of a combination of
bending and shearing\cite{Sokolnikoff_book} as well. We define here
the strain as the most straightforward interpolation of the two cases
that recovers the short cylinder result for $s\to0$ and the long
cylinder form in the limit $s\to\infty$:
\begin{align}
\ve_{s,\mathrm{max}}&=
\frac{X}{H(1+\frac{H}{8R})}=
\frac{(f\pi)^{1/3}X}{2s^{2/3}N^{1/3}\bar d(1+s/4)}\cr
\bar\ve_s&=\frac{4 \ve_{t,\mathrm{max}}}{3\pi},
\end{align}
Applying Hooke's law to the whole strain field yields the
interpolating stress
\begin{align}
\sigma_{s,\mathrm{max}}&=
\frac{F_x\left(2\nu+2+\frac{H^2}{3R^2}\right)}{\pi R^2(1+\frac{H}{8R})}=\cr
&=\left(\frac{fs}{\sqrt{\pi}N}\right)^{2/3}
\frac{\left(2\nu+2+\frac{4s^2}{3}\right)}{(1+s/4)}
\frac{F_x}{\bar d^2}
\cr
\bar\sigma_s&=\frac{4 \sigma_{s,\mathrm{max}}}{3\pi}
\quad.
\end{align}

\begin{figure}
\includegraphics[width=\columnwidth]{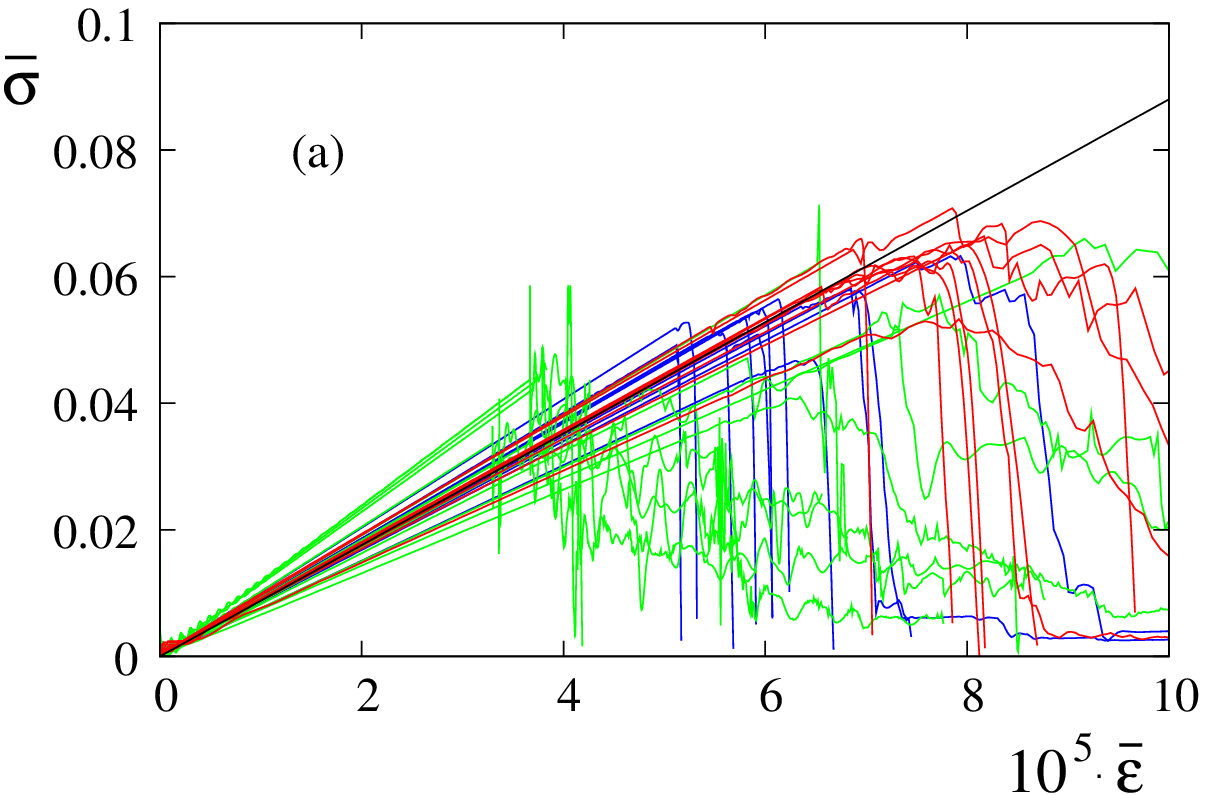}
\includegraphics[width=\columnwidth]{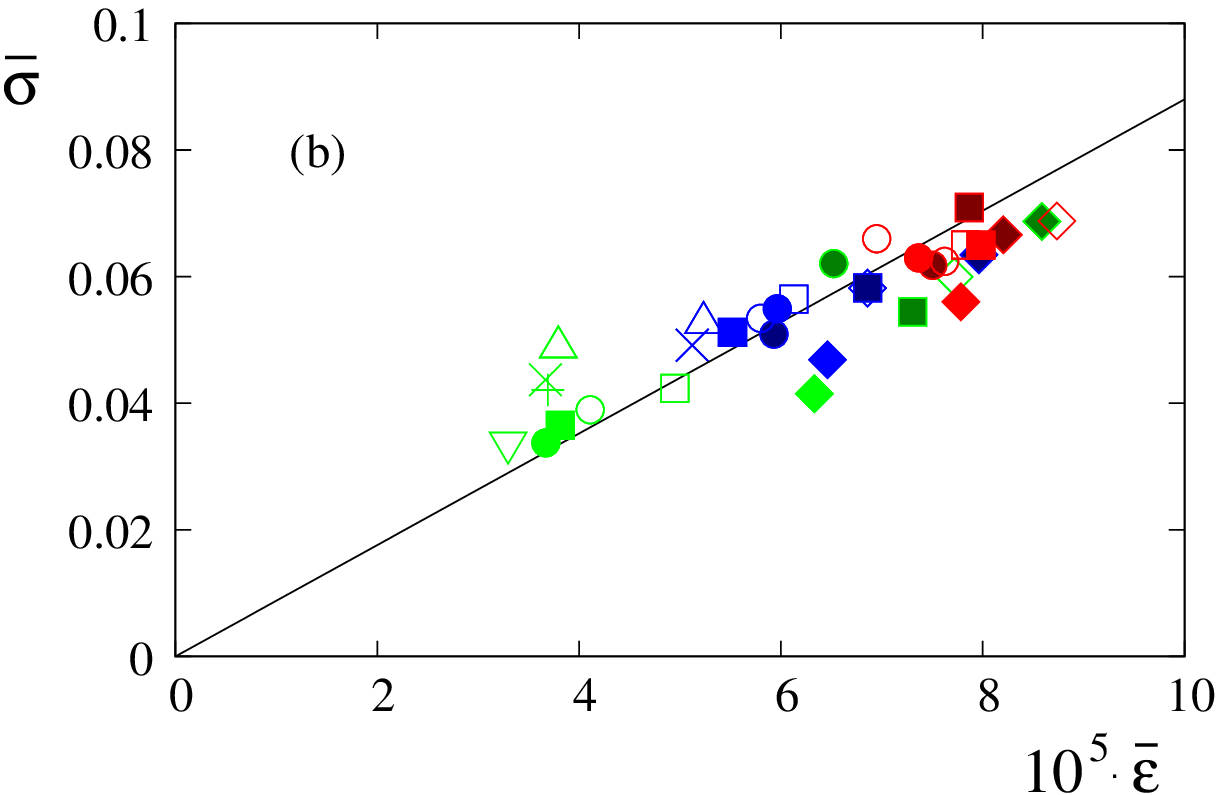}
\includegraphics[width=\columnwidth]{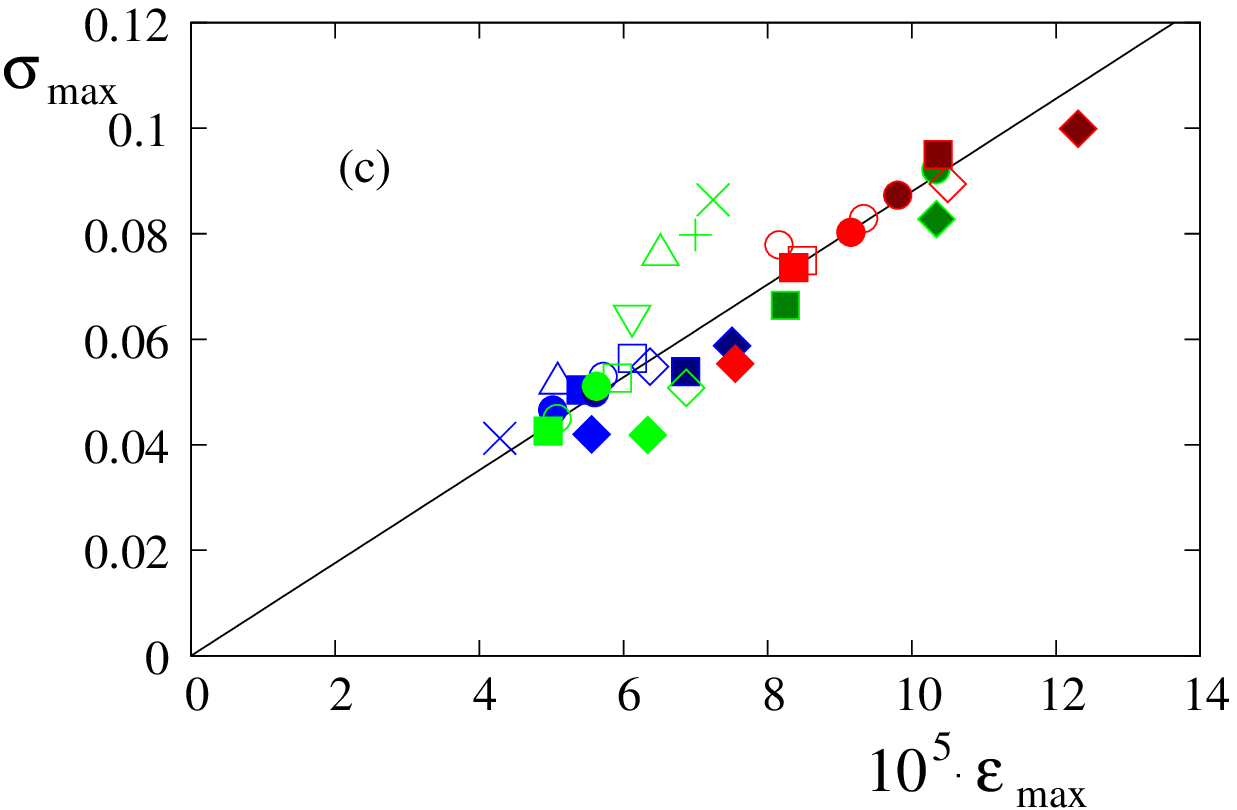}
\caption{\label{Fig_scaling} Color online. (a) Stress strain evolution
(b) The maximum of the average stress, (c) End of elasticity
(first broken contact) in the stress/strain plain. Colors represent
experiment type {\em pull}: blue, {\em shear}: green, {\em twist}:
red. Symbol fill intensity indicates system size $N$: $200$: half,
$2000$: empty, $10000$: full. Symbol types show aspect ratio $s$: 1:
$\diamond$, 2: $\square$, 3: $\bigcirc$, 5: $\bigtriangleup$, 7:
$\bigtriangledown$, 8: $+$, 10: $\times$
}
\end{figure}

Figure \ref{Fig_scaling} (a) shows the average stress/strain curves
for different experiments, system size, sample length. The measured
Young's modulus varies around a mean value $E{\simeq}880$, which
is 15\% less than the theoretical value of
(\ref{Eq_youngsestimate}). There is a slight scatter of the linear
slope of different systems but in spite of the moderate system size
still one can observe a convincing data collapse with different system
size, aspect ratio and test type.

In Fig.~\ref{Fig_scaling} (b) the maxima of the average stress/strain curves
are shown as function of the average strain. It is evident that the
elastic behavior is very close to the Hooke's law and can be
described by a single Young's modulus. The average macroscopic stress
and strain are always smaller than the scale
(\ref{Eq_predict}), as must be expected due to spatial disorder.

The maximal stress/strain values at the moment of the first broken
contact are plotted in Fig.~\ref{Fig_scaling} (c). The values are
scattered around $\ve_*$ of (\ref{Eq_predict}). We note two important
features: (i) points which are above $\ve_*$ are mainly for system
size $N{=}200$ and for shear and twist tests. This is the result of
the inhomogeneities of our system. Due to the preparation procedure
the volume fraction at the lateral surface of the cylinder is larger
than in the bulk. This was tested by repeating a few tests with a sample
cut out from a bigger one, and this feature disappeared. (ii) Maximal
stresses for long cylinders in shear tests are somewhat larger than
expected, these are the result of the same skin effect during the
bending of the sample.

\subsection{Yield criteria}
\label{Sec_yieldcriterions}

\begin{figure}
\includegraphics[width=\columnwidth]{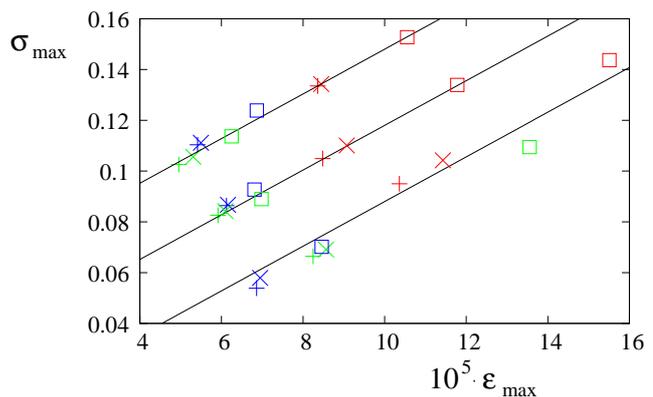}
\caption{\label{Fig_criterioncompare} The stress and the strain points
where the first contact was broken. Colors represent experiment type
{\em pull}: blue, {\em shear}: green, {\em twist}: red. Symbols
indicate the different break criterions: Full $+$, Simplified:
$\times$, Decoupled $\square$. System size $N{=}200$, 2000, $10000$,
are from bottom to top respectively with artificial shifting of $0.3$.
The aspect ratio was chosen to be $s{=}2$
}
\end{figure}

In the simulations we used all three criterions introduced in
Sec.~\ref{Sec_breakcrit}. We performed test using exactly the same initial
configuration with the different break criterions.
Fig.~\ref{Fig_criterioncompare} shows the summary of the results.

The results show hardly any difference between the two coupled
criterions (simplified (\ref{eq_lesscrude}) and full
(\ref{eq_Phi2}), however the decoupled criterion allows the system
to evolve without failure beyond the point where the others already
broke.

It is remarkable that the simple coupling of the different deformation
modes leads to indistinguishable failure points for large systems. For
small systems the only difference is in the twist deformation.

The elapsed time during simulations was also measured. We observed no
measurable difference in simulation duration of tests with decoupled
and simplified criterions. Simulations with the full criterion on
average require 55\% more time independently of system size. This
advocates the use of the simplified criterion if computational
time is critical but this overhead is relatively small in spite of
the numerical maximization of (\ref{eq_Phi2}) which is in general
affordable if precision is required.

\subsection{Failure process}
\label{Sec_failuretype}

\begin{figure*}
\includegraphics[width=0.8\columnwidth]{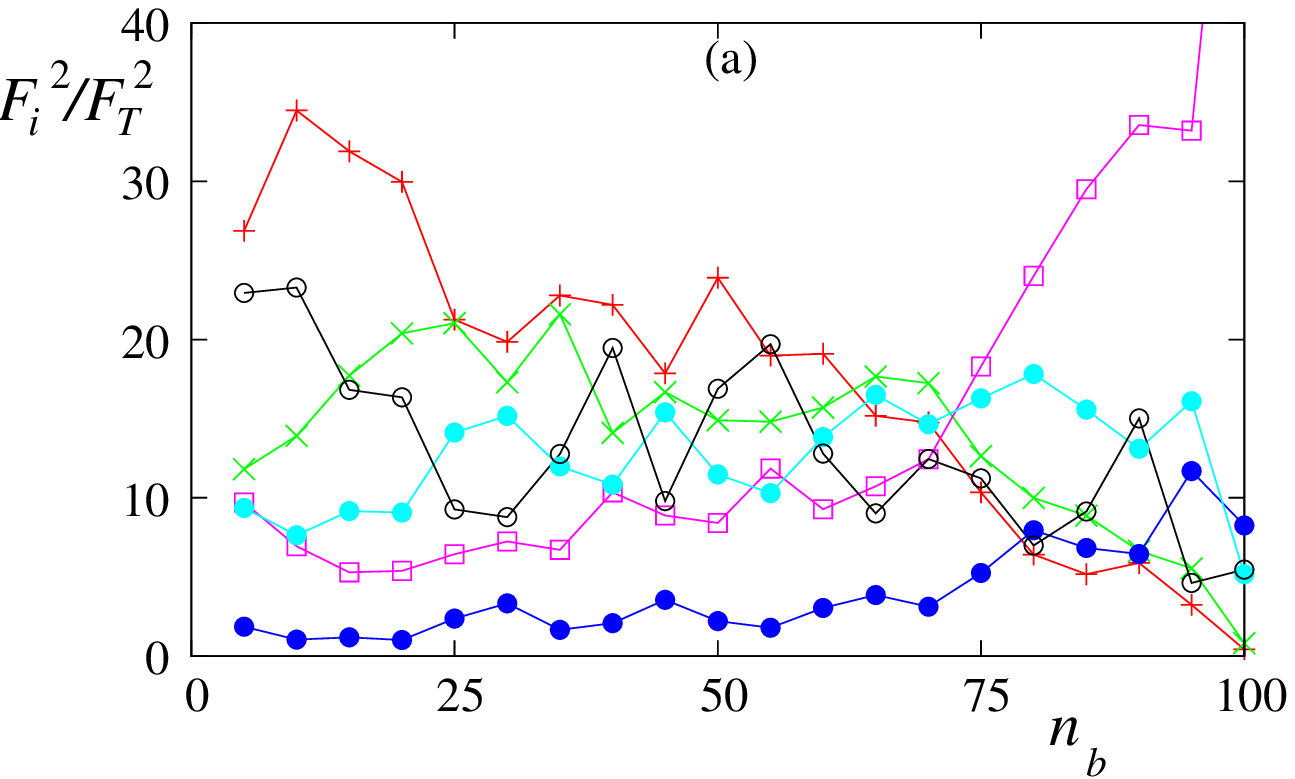}
\includegraphics[width=0.2\columnwidth]{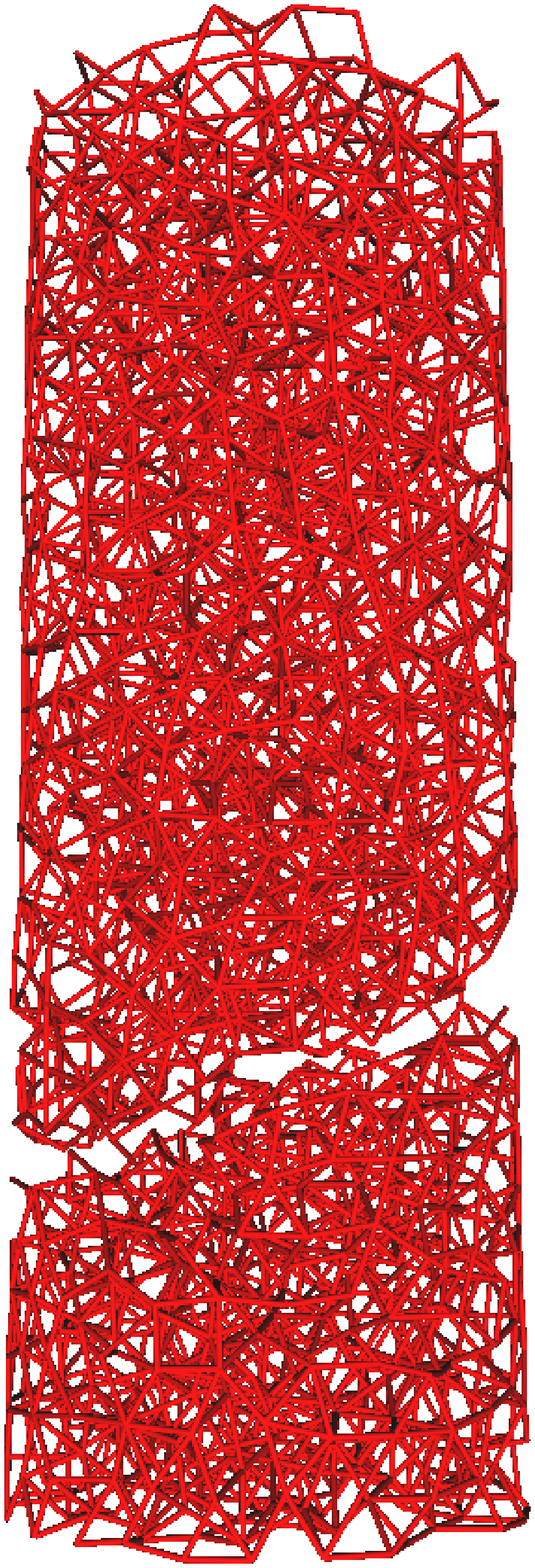}
\includegraphics[width=0.8\columnwidth]{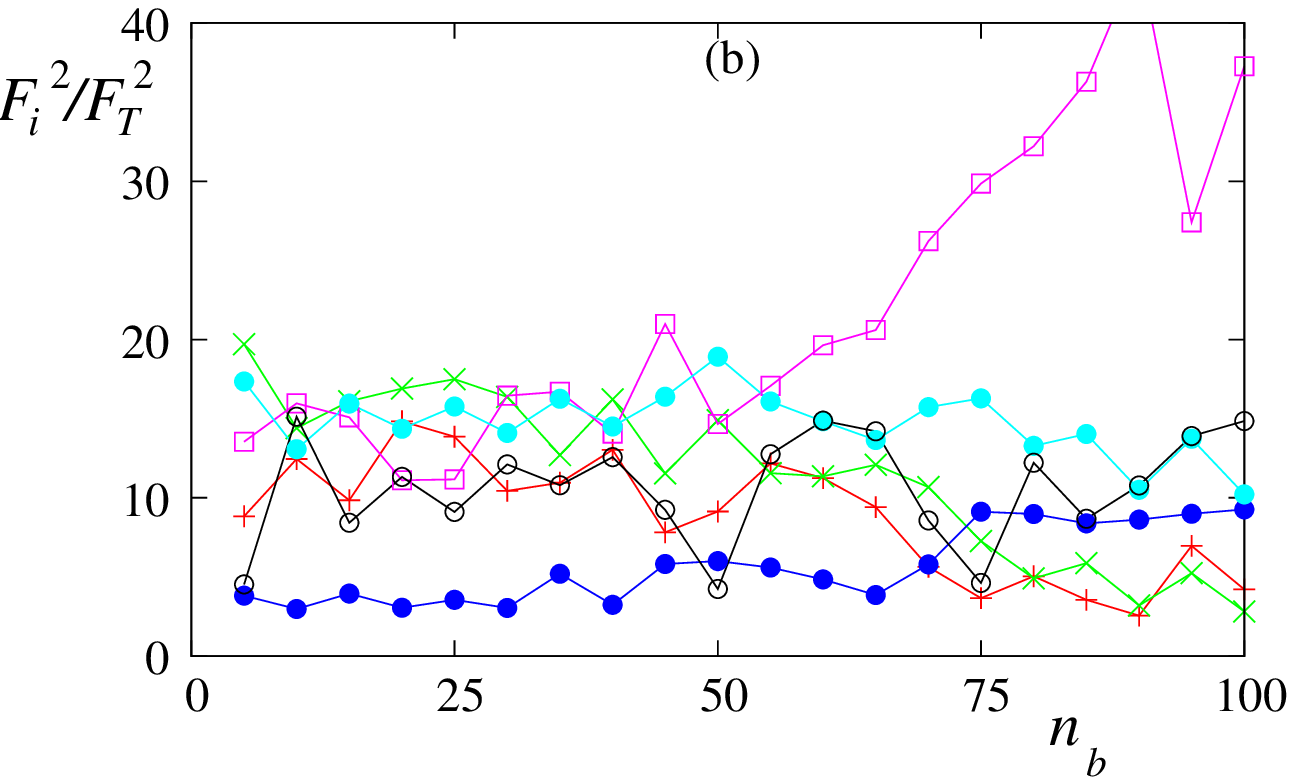}
\includegraphics[width=0.2\columnwidth]{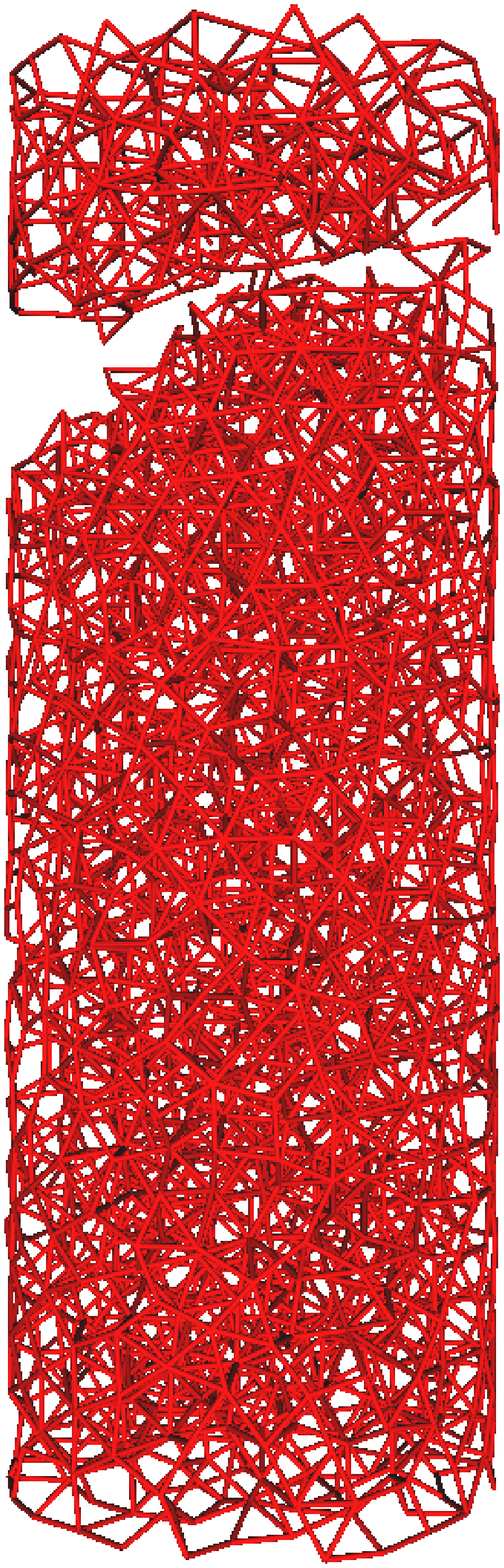}
\newline
\includegraphics[width=0.8\columnwidth]{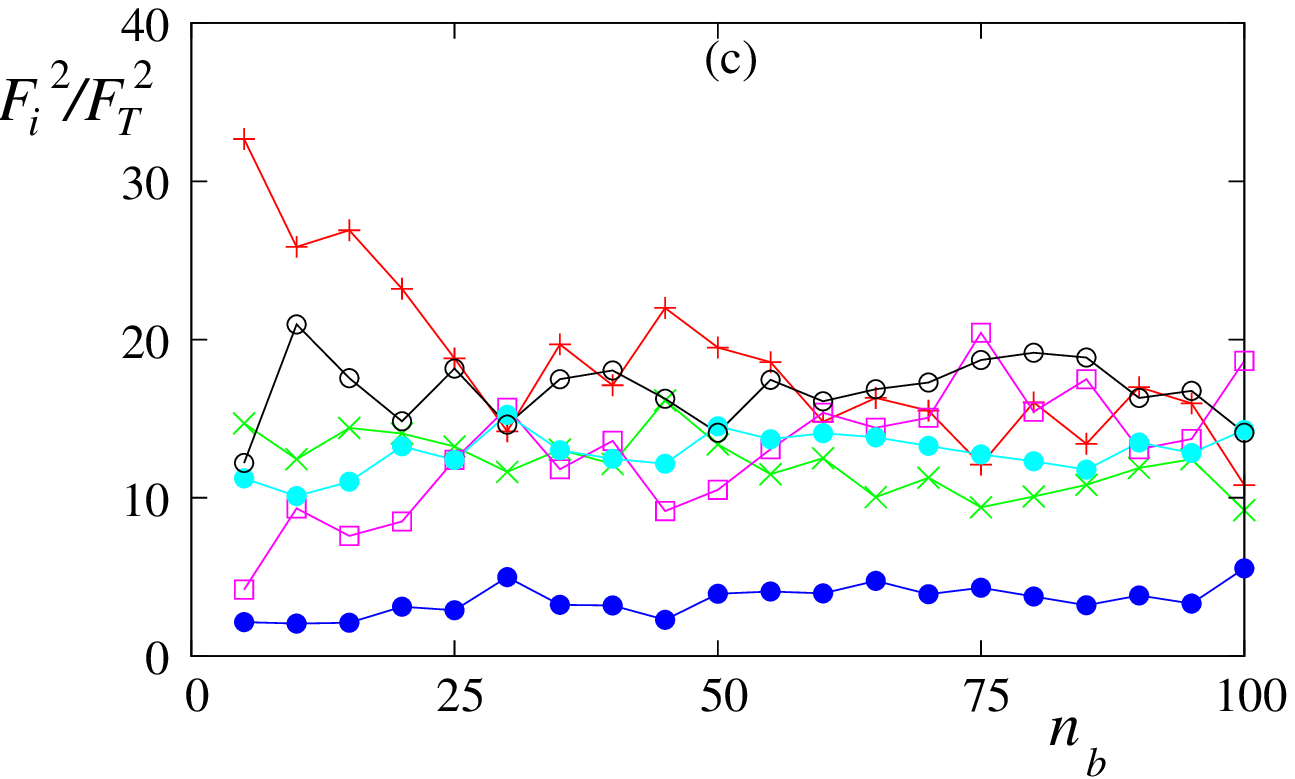}
\includegraphics[width=0.2\columnwidth]{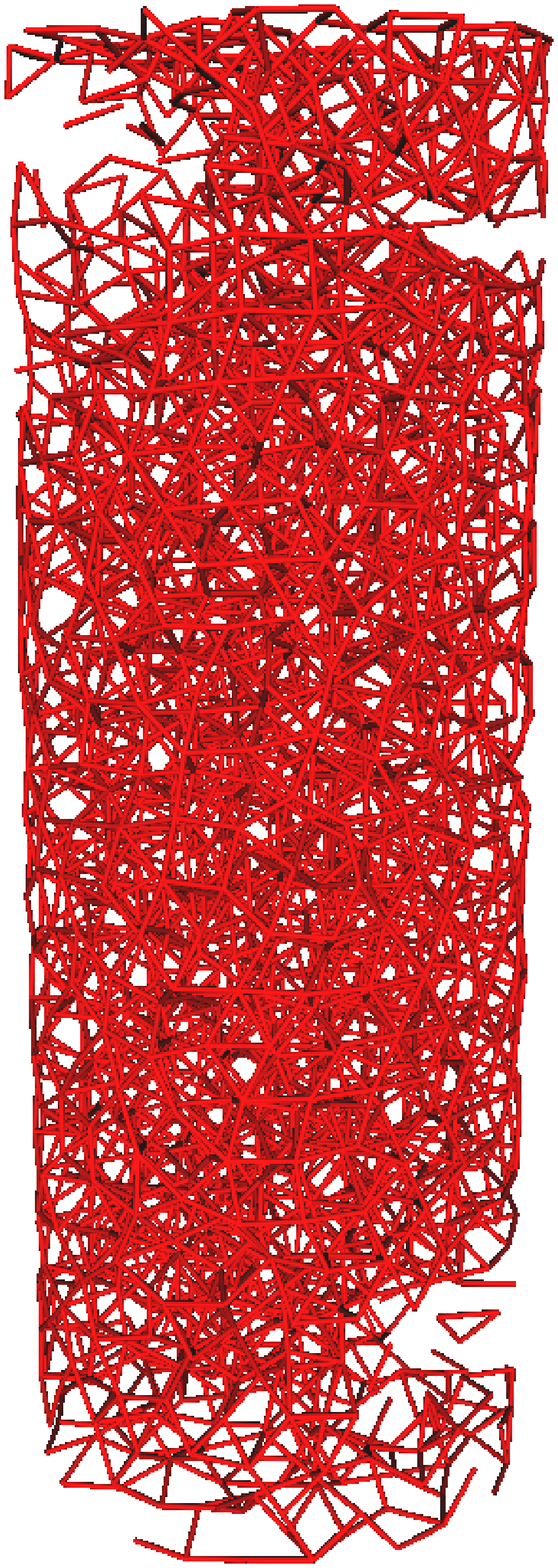}
\includegraphics[width=0.8\columnwidth]{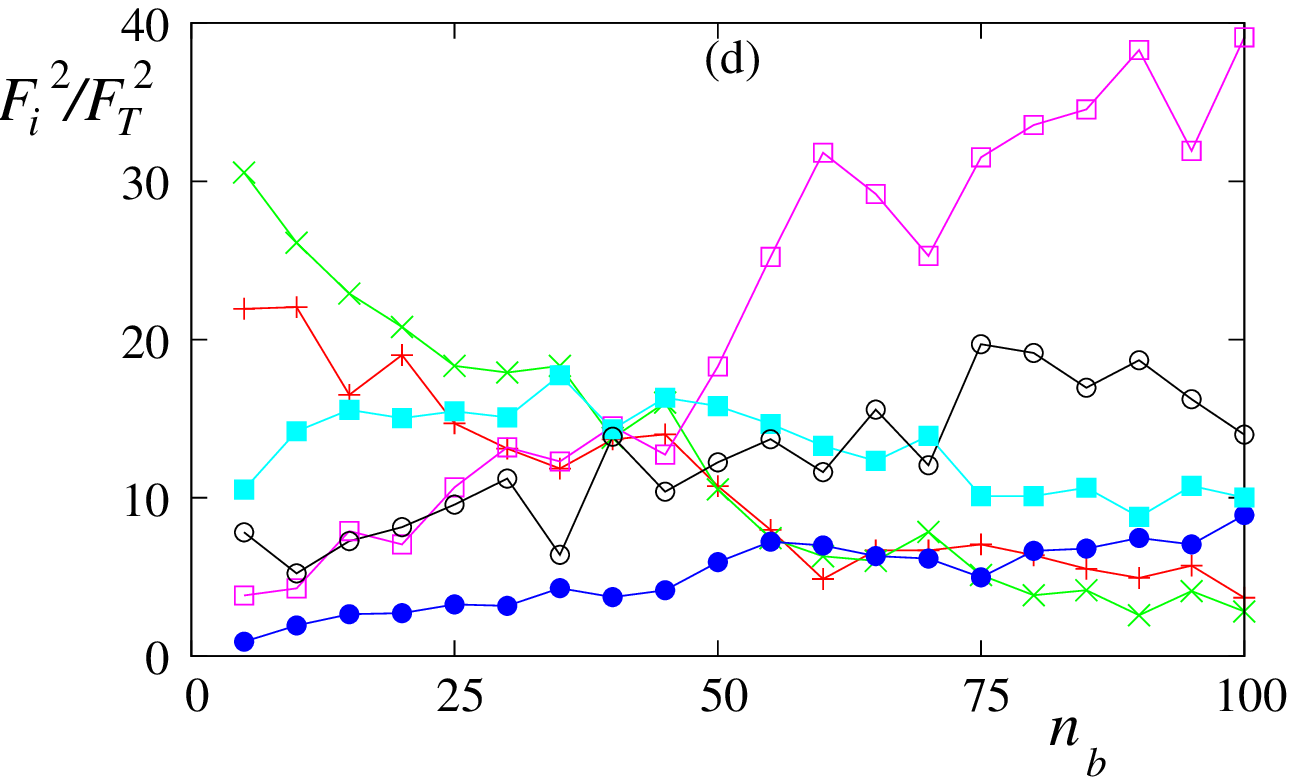}
\includegraphics[width=0.2\columnwidth]{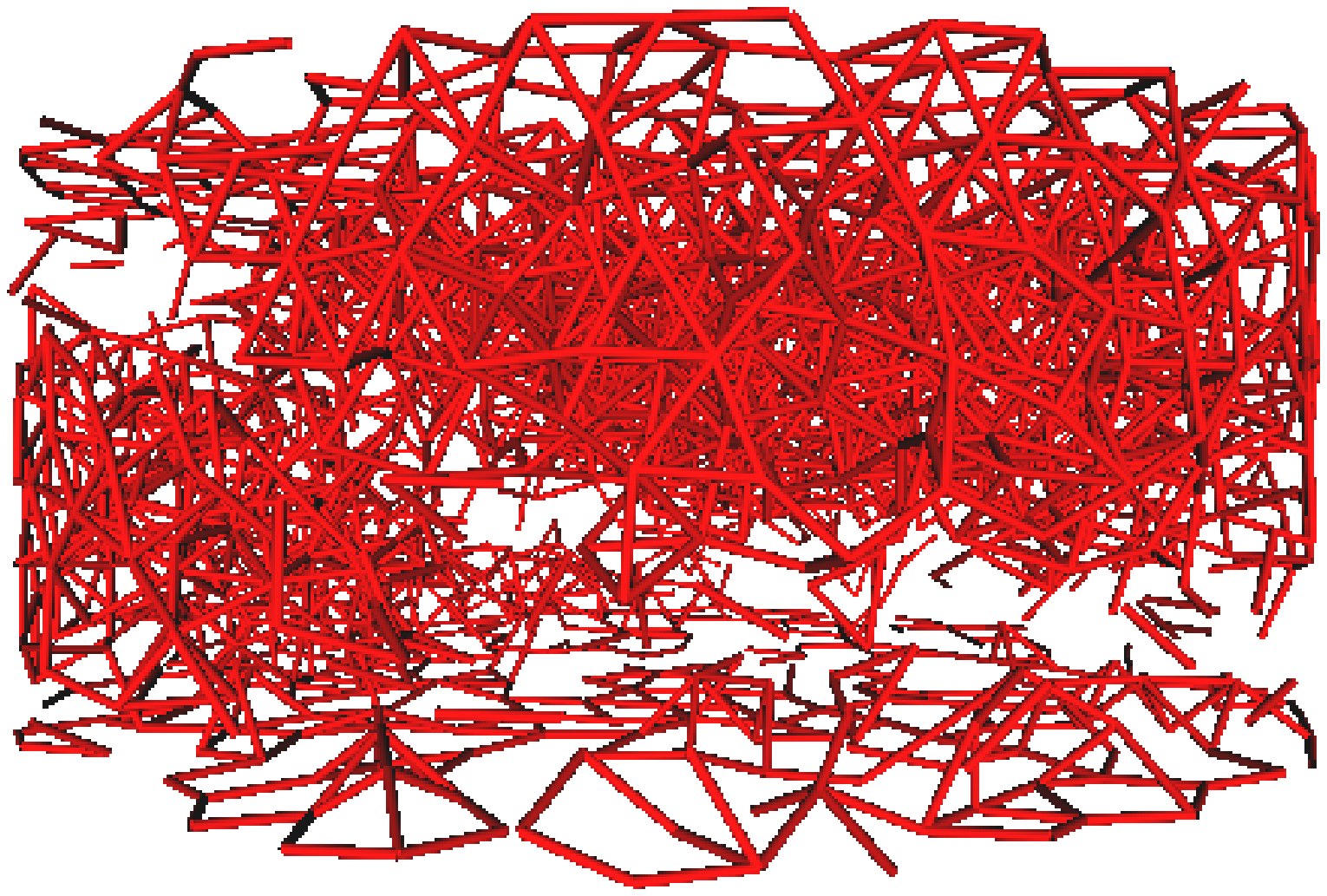}
\caption{\label{Fig_close} Color online. Average (sample and time)
deformation forces in just broken contact bridges scaled by the
maximal force versus $\tilde N_b$. (a) Pull test with $s{=}3$ (b)
Twist test with $s{=}3$ (c) Shear test with $s{=}3$, (d) Shear test
with $s{=}0.5$. The symbols represent the following deformation
component: tensile: $+$, shear: $\times$, tilt: $\bullet$, torsion:
$\square$, tilt+tensile: $\blacksquare$, torsion+shear: $\bigcirc$.
On the right hand side of the plots a typical example of force network
plot is shown for the given system after the test.
}
\end{figure*}

In this section we study the process of the failure in the tests.
Our simulations allow the study of each individual broken contact.
The question we want to answer here is: Which is the deformation type
that is the most relevant for a give test.

We identify 6 different deformation types which are the core of the
simplified break criterion (\ref{eq_lesscrude}):
\begin{eqnarray}
\mathrm{tensile:} &\quad& F_z^2/4\cr
\mathrm{shear:} &\quad& F_x^2+F_y^2\cr
\mathrm{tilt:} &\quad& 4T^2_x/a^2 \cr
\mathrm{torsion:} &\quad& 4T^2_z/a^2\cr
\mathrm{tilt+tensile} &\quad& 2|F_zt_x-2F_xT_z|/a\cr
\mathrm{torsion+shear} &\quad& 4|F_yT_z|/a,
\end{eqnarray}

In order to study the relevance of the above deformation types, the
following procedure was used: At the moment of a contact break, the
above quantities are recorded.

Since we have a random packing, the values show huge variations from
contact to contact. An average is necessary to be able to observe the
different trends. We averaged the recorded limit forces for each
successive 5\% of broken contacts. So if $N_b(t)$ is the number of
broken contacts at time $t$ then the $x$ axes shows
$n_b{=}N_b(t)/N_b(T)$, with $T$ being the time at the end of the
experiment. This representation allows us to average also for
different sample realizations (5-10) so that early and late stages of
failure can be identified, as shown in Fig.~\ref{Fig_close}.

The beginning of the pull test [Fig.~\ref{Fig_close} (a)] is marked by
contacts strained in the tensile direction. This dominance is clear
for the first 25\% broken contacts but remains the most important up
to 60\% broken contacts.  The end stage (after 75\% of the contacts
were broken), is surprisingly characterized by contacts breaking due
to torsion. Our view is that this is due to breaking of small chains,
namely a single particle having two contacts one to each side of the
sample. In general, since all ``strong'' contacts are already broken,
its two contacts are loaded mainly with a combination of torsion and
tilting. A remarkable feature is that tilting gets stabilized by shear
and tensile forces, while torsion is not. This can be seen by the fact
that tilting+tensile deformation is almost the second most important
loading at late stages. Since we have used a rather narrow contact
diameter $a{=}0.1\bar d$, contacts break easily under torsional load.

Twist tests [Fig.~\ref{Fig_close} (b)] show a surprisingly similar
picture to pull experiment, except that at early stages it is the
shear deformation which dominates. The late stage is also dominated by
torsional deformation.

The behavior of shear tests are fundamentally different depending on
the aspect ratio. If the aspect ratio is small [$s<1$,
Fig.~\ref{Fig_close} (d)] the overall behavior is similar to the twist
deformation: shear dominates at early stages and torsion at late ones.

\begin{figure}
\includegraphics[width=\columnwidth]{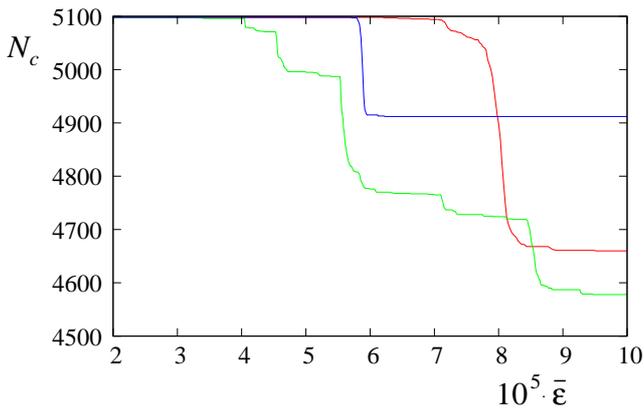}
\caption{\label{Fig_contactnumber} (Color online) The number of
contacts as function of the average strain for $s{=}3$ and $N{=}2000$,
for the tests: {\em pull}: blue, {\em shear}: green, {\em twist}: red
}
\end{figure}

It is common to all of the above cases that the stress/strain
behavior of the sample is determined by the contacts that broke in
the early stages, moreover the whole fracture is realized very fast
in a cascade like manner as shown on Fig.~\ref{Fig_contactnumber}, as
to be expected in the brittle limit.

The only test where a different scenario can be observed is the shear
test with long cylinders [Fig.~\ref{Fig_close} (c)]. At early stages the
tensile deformation dominates. This is due to the bending of the
sample. The force network shows that the contacts start to break at
two points simultaneously.  These points are the ones where the
material is stretched the most (right at the moving ``front'' of the
driving layer).  The stress is very localized so despite the damage
the sample has still a significant resistance which is manifested by
the long oscillating tail of the stress/strain curve
[Fig.~\ref{Fig_scaling} (a)] and the slow step-wise decreasing of the
number of contacts.


\section{Conclusion}
\label{Sec_conclusion}
In this paper we introduced a new model for DEM simulations with
breakable cemented contacts. Contacts were considered as small, flat,
elastic cylinders which keep the particles in contact until the shear
stress in the cementing cylinder reaches a maximal value conforming to
the Tresca criterion of failure. This allowed for deriving the elastic
and the failure behavior of the contact on a common basis. In the
limit of very flat cylinders, the stiffness matrix (\ref{eq_mstiff})
gets decoupled, justifying this common a priori modelling Ansatz.


The problem of finding the maximal shear stress in an arbitrarily
loaded cylinder turned out to possess analytic solutions too
complicated for practical purposes (but is numerically well
accessible). Therefore we considered two approximations: The
\emph{decoupled criterion} coincides with the exact maximization only
for single loading modes, while the \emph{simplified criterion} agrees
with the latter in more general situations, due to taking into account
load combinations. We showed that decoupling the loads leads to a
rather large error of 25\%, whereas the simplified criterion agreed
with the full one within statistical errors. This applies to our
tests, but in the case of the full criterion turning out to be
necessary in other situations, the found overhead of 55\% for the
numerical maximization is affordable.

As a benchmark, we have done pull, shear and torsion tests with
cylindrical specimens created from glued particles. These exhibited,
up to the failure point, an elastic behavior with the same Young's
modulus for all experiments and test parameters. This enabled us to
collapse all different test results onto one single curve. The
deviations of the theoretical prediction for the specimens' elastic
properties were below 15\%.

During the tests we payed special attention to the dominant loading of
the contacts that are about to break. We found that the first half of
broken contacts in pull tests mostly break due to tensile stress, in
twist tests due to shear stress, and in shear tests of very short
samples due to shear stress as well. Long samples behave differently
in shear tests, though, due to bending the dominating load gets
tensile instead of shear. Finally, the late stages of the failure is
mainly characterized by contacts breaking under torsional load due to
break up of twisted weak chains connecting the almost separated parts.

For the sake of a clearer study of this relative importance of the
loading types and the different failure criteria, we used a constant
diameter of the contact cylinders. In future works, the stochastic
nature of the geometry and strength of the contact, possibly obeying a
wide distribution\cite{Birkenfeld}, will be taken into account.


\section{Acknowledgements}
The authors acknowledge the financial support from the German Research
Foundation (DFG grant BR 3729/1) and thank Sandia National
Laboratories (of the US Department of Energy) for putting LAMMPS under
the GNU Public License.

\bibliographystyle{spmpsci}
\bibliography{lit}

\end{document}